\documentclass{article}
\usepackage{spconf,amsmath,graphicx,algorithm,algorithmic,multirow,cite,tabularx}
\usepackage{color}

\title{NDVQ: Robust Neural Audio Codec with Normal Distribution-Based Vector Quantization}

\name{{\centering
Zhikang Niu$^{1,2*}$, Sanyuan Chen$^2$, Long Zhou$^2$, Ziyang Ma$^1$,
Xie Chen$^1$, Shujie Liu$^2$ \thanks{$*$Work done during internship at Microsoft Research Asia.} }}

\address{$^1$MoE Key Lab of Artificial Intelligence, AI Institute, X-LANCE Lab, \\Shanghai Jiao Tong University, China
    $^2$Microsoft Research Asia, China}

\begin{document}
\ninept
\maketitle
\begin{abstract}
Built upon vector quantization (VQ), discrete audio codec models have achieved great success in audio compression and auto-regressive audio generation. However, existing models face substantial challenges in perceptual quality and signal distortion, especially when operating in extremely low bandwidth, rooted in the sensitivity of the VQ codebook to noise. This degradation poses significant challenges for several downstream tasks, such as codec-based speech synthesis.
To address this issue, we propose a novel VQ method, Normal Distribution-based Vector Quantization (NDVQ), by introducing an explicit margin between the VQ codes via learning a variance. Specifically, our approach involves mapping the waveform to a latent space and quantizing it by selecting the most likely normal distribution, with each codebook entry representing a unique normal distribution defined by its mean and variance. Using these distribution-based VQ codec codes, a decoder reconstructs the input waveform. NDVQ is trained with additional distribution-related losses, alongside reconstruction and discrimination losses. 
Experiments demonstrate that NDVQ outperforms existing audio compression baselines, such as EnCodec, in terms of audio quality and zero-shot TTS, particularly in very low bandwidth scenarios.
\end{abstract}
\begin{keywords}
Audio Compression, Audio Codec, Normal Distribution Vector Quantization, Text to Speech
\end{keywords}
\section{Introduction}
\label{sec:intro}

Recently, the audio compression task has garnered increasing interest due to the explosion of streaming media technology and the advancement of generative models. Audio codec research is fundamentally divided into two paradigms: traditional handcrafted algorithms and neural network-based models. The former, including Modified Discrete Cosine Transform (MDCT) \cite{mdct} and Linear Predictive Coding (LPC) \cite{lpc}, has been the backbone of leading traditional codecs such as Opus \cite{opus} and Enhanced Voice Service (EVS) \cite{evs}. They are prized for their efficiency and versatility but often suffer from significant information loss, especially for the scenarios of low bitrates.

On the other hand, neural network-based models have achieved significant breakthroughs \cite{soundstream,encodec,dac} by leveraging residual vector quantization (RVQ) and generative adversarial training. 
SoundStream\cite{soundstream} is one of the first universal neural audio codecs, supporting varying bitrates and diverse audio types. It comprises a SEANet encoder, a SEANet decoder, and RVQ layers. EnCodec\cite{encodec} builds upon the SoundStream framework and architecture, incorporating several modifications. EnCodec utilizes a multi-scale STFT discriminator to enhance audio quality and a loss balancing mechanism to train stability. Moreover, EnCodec employs a Transformer-based language model to accelerate the compression and decompression process. Despite these advancements, both of them struggle with low codebook utilization and codebook collapse. This is attributed to the fact that previous vector quantization-based codec models quantize the vectors from the encoder based on nearest neighbour selection with Euclidean distance, failing to accommodate small variations, especially for the scenarios at low bandwidths and the input with noises,  adversely affecting tasks like codec-based speech synthesis. To address these problems, Descript-audio-codec (DAC)\cite{dac} introduces significant design alterations, including a periodic activation function, a multi-scale mel loss, and factorized, L2-normalized vector quantization. While DAC achieves state-of-the-art performance in audio coding and supports a variety of sample rates, its parameter count is fourfold that of EnCodec.

To address codebook collapse, improve robustness, and maintain a low parameter count, we introduce the Normal Distribution Vector Quantization (NDVQ) method, which explicitly builds a safety margin by incorporating learnable variance in the codebook. Unlike traditional VQ methods that cluster hidden states with codes in the codebook based on Euclidean distance, NDVQ represents each code in the codebook as a normal distribution, and the hidden states are clustered into different codes following different distributions. 
During training, NDVQ ensures that each code is as distinct as possible in the latent space, preventing codebook collapse and low codebook utilization. By fully utilizing the codebook, NDVQ effectively increases its expressive capacity, making the model more suitable for audio auto-regressive tasks. NDVQ further enhances robustness by sampling vectors from the corresponding distribution for audio reconstruction during training.

Our comprehensive experimental evaluations and ablation studies demonstrate the effectiveness of the NDVQ approach in audio compression and codec-based speech synthesis, particularly at low bitrates where maintaining audio quality is more challenging. The more distributed representation of NDVQ significantly bolsters its generalization capability, enabling the encoder to better adapt to diverse data in unseen scenarios. Trained with speech data in one language (English) and one domain  (LibriTTS of audiobook data), NDVQ 
can reconstruct audio in high quality in other languages (Chinese) and domains (VCTK dataset with various accents). Further experiments with codec-based speech synthesis confirm that NDVQ improves the generative model's performance, suggesting its potential as a universal audio codec in future developments.
\begin{figure*}[t]
    \centering
    \includegraphics[width=0.9\linewidth]{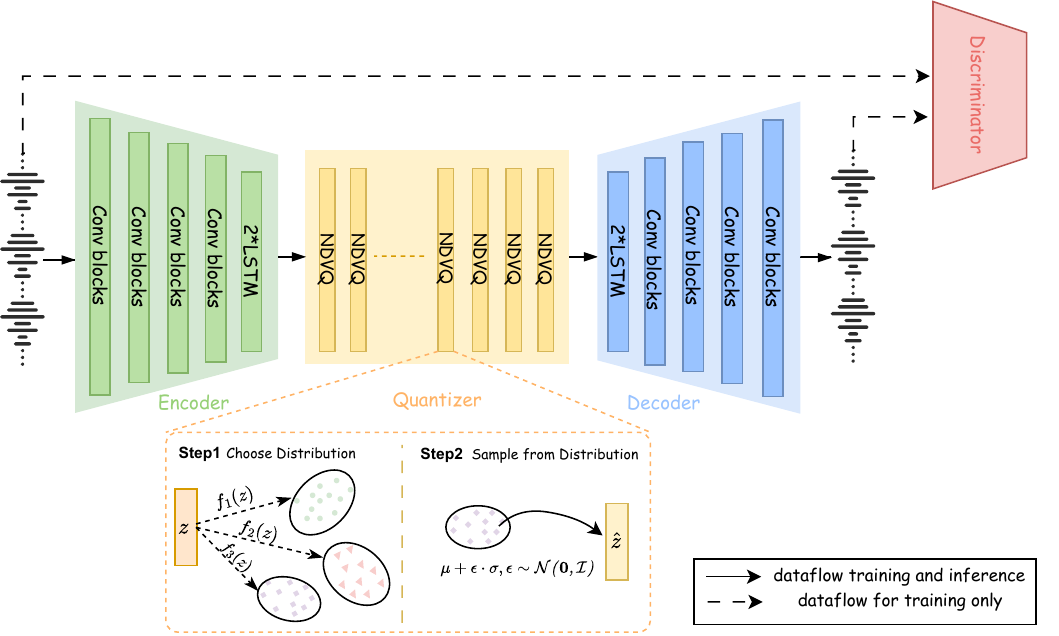}
    \caption{NDVQ: an encoder-decoder based robust neural audio codec that utilizes normal distribution-based vector quantization. NDVQ selects the most similar probability distribution to quantize the encoder output and employs a re-parameterization trick to obtain quantization results from the codebook data mean and variance. The discriminator is utilized only during training.}
    \label{fig:normal vector quantization pipeline}
\end{figure*}
Overall, the main contributions include: 
\begin{itemize}
    \item A novel NDVQ method representing each codebook entry as a normal distribution instead of deterministic multi-dimensional vectors.
    \item An extensive comparative analysis with the state-of-the-art EnCodec demonstrating superior performance, particularly in low-bandwidth scenarios.
    \item A comprehensive set of ablation studies validating the significance of components in our proposed NDVQ model.
\end{itemize}

\section{Methods}
\label{sec:methods}

Due to the susceptibility to noise, codebook collapse, and low codebook usage, audio compression with high fidelity and low bitrate is still a challenging task for neural codec modeling. To deal with these problems, in this section, we introduce our new codec model NDVQ. For a given audio $x\in [-1,1]^{T}$, where $T$ is the total number of samples, calculated as $T = d \times sr$ with duration $d$ and sample rate $sr$, our objective is to compress and transmit this audio information using the fewest possible bits without compromising quality.

As shown in Figure \ref{fig:normal vector quantization pipeline}, our proposed model, inspired by the EnCodec framework \cite{encodec}, consists of four key components: an encoder, a novel residual normal distribution vector quantizer (NDVQ), a decoder, and a discriminator. The process begins with the encoder $\rm{Enc}$ mapping the input waveform $x$ into a latent representation $z(x)$. The latent representation is then processed by our NDVQ, which produces quantized latent codes by sampling from a normal distribution, resulting in the quantized representation $\hat{z(x)}$. The decoder $\rm{Dec}$ uses these quantized results $\hat{z(x)}$ to reconstruct the original audio, yielding $\hat{x}$, while the discriminator $D$ is employed to enhance the quality of the reconstructed audio. 

\subsection{Encoder \& Decoder}
Drawing on the insights from prior research  \cite{soundstream,encodec,dac}, we choose SEANet\cite{seanet} without skip connections as our encoder and decoder. This symmetric network comprises four convolutional blocks and a sequence of two stacked long short-term memory networks (LSTMs). 

The encoder $\rm{Enc}$ adopts down-sampling strides of $(2,4,5,8)$, a 1-second 24khz audio can be compressed by a factor of $2\times4\times5\times8=320$, resulting in 75 frames. The decoder mirrors this encoder down-sampling strides, up-sampling in the reverse order to reconstruct the audio.

\subsection{Residual Normal Distribution Vector Quantization}

Building upon the advancements in neural audio compression \cite{soundstream,encodec,dac}, our approach integrates the strengths of residual vector quantization (RVQ) with the robustness of probabilistic models. Traditional VQ methods exhibit two main weaknesses: they tend to have a low codebook usage and are highly sensitive to noise because of deterministic nearest neighbor selection, which fails to accommodate minor variations in the input data, leading to poor reconstruction quality and background artifacts in the non-reconstruction tasks, such as codec-based speech synthesis.

To mitigate these issues, we redefine the quantization problem from deterministic nearest neighbor selection to a probability distribution selection problem. Our Normal Distribution Vector Quantization (NDVQ) codebook leverages the normal distributions which consist of two parts: data mean $\mu$ and variance $\sigma$. During the training, we use the probability density function to calculate which probability distribution the latent representation belongs to, and sample from this distribution to obtain the quantized result through the re-parameterization trick by data mean and variance. NDVQ has 32 codebooks and each has 1024 codes (distributions) following the EnCodec framework \cite{encodec}.

The NDVQ quantization process involves sampling from the learned distributions. For a given latent representation $z(x)$, the corresponding distribution for quantization is identified using the probability density function: 
\begin{equation}
\resizebox{.91\hsize}{!}{$ 
    f(z(x)) =\operatorname*{argmax} \sum_{i=1}^{D} \left(-\frac{1}{2}\left(\frac{z(x)_i-\mu_i}{\sigma_i}\right)^2\right) - \log (\sigma_i \sqrt{2\pi}),
    $}
\end{equation}
where $\mu_i$ and $\sigma_i$ are the mean and variance of the normal distribution for the $i^{\text{th}}$code, and $D$ is the codebook dimension. 

\begin{algorithm}[ht] 
\caption{Vector Quantization of NDVQ in Training}  
\begin{algorithmic}  
\REQUIRE $z(x) = \text{enc}(x)$ the output of the encoder, \\ \quad \quad \quad vector quantizers $Q_i$ for $i = 1..N_q$. 
\ENSURE the quantized $\hat{y}$  
\STATE $\hat{y} \gets 0.0, \hat{y}_i \gets 0.0$  
\STATE $\text{residual} \gets y$  
\FOR{$i = 1$ to $N_q$}  
    {\color{blue}
    \STATE $\mu(x),\sigma(x) \gets Q_i(\text{residual})$
    \STATE $\hat{y}_i \gets \mu(x) + \epsilon \cdot \sigma(x), \epsilon \sim \mathcal{N(\mathbf{0, I})}$
    }
    \STATE $\hat{y} \gets \hat{y} + \hat{y}_i $
    \STATE $\text{residual} \gets \text{residual} - \hat{y}_i$  
\ENDFOR  
\RETURN $\hat{y}$  
\label{rndvq}
\end{algorithmic}  
\end{algorithm} 

As shown in Algorithm 1, the model selects the most appropriate distribution for $z(x)$ by evaluating its probability density across all codebook entries. Given the selected distribution, we apply the re-parameterization trick \cite{kingma2013auto} to sample a quantized representation $\hat{z}(x)$, which enhances the model's robustness to input perturbations by introducing a stochastic element via the noise variable $\epsilon \sim \mathcal{N(\mathbf{0}, I)}$. During inference, we only use the mean part $\mu$ as the quantization results.

Instead of leveraging only one vector (with deterministic nearest neighbor selection) to represent the quantized results during model training, NDVQ represents each codebook's code as a normal distribution (with a mean and a variance). The introduced variance forces each code to be as far away as possible in the latent space so it can not only improve the robustness of the code selection but also effectively enlarge the codebook representation space capacity compared to the same codebook size VQ methods. With the introduced variance and random sampling, our NDVQ can better fit the audio data with a better capability to deal with the noises, codebook collapse, and poor codebook usage problems, to improve the downstream tasks such as codec-based speech synthesis.

\subsection{Discriminator}
The advancement of generative models, notably GANs\cite{gan}, diffusion networks\cite{ddpm}, and auto-regressive networks\cite{gpt2}, has been a driving force in speech synthesis, yielding high-fidelity audio outputs. Among them, GAN models are particularly notable for their ability to produce high-fidelity audio at high inference speeds.

we adopt the multi-scale STFT-based (MS-STFT) discriminator following EnCodec\cite{encodec}. This discriminator consists of several STFT-based sub-discriminators, each designed to analyze audio at varying spectral resolutions. The sub-discriminators are tailored to specific scales, employing STFTs with different receptive fields and hop lengths, and five distinct scales with STFT window lengths set to $[2048,1024$, $512,256,128]$ are used to model the signal comprehensively. With these 2D convolutional layers, each layer of the sub-discriminators outputs a logit indicating the authenticity of the signal, and all these logits are used to get the final loss using a relative feature-matching loss.
\begin{table*}[h]  
    \caption{Performance Comparison of Audio Codec Models at Various Bitrates}  
    \centering  
    \label{main_table}  
    \begin{tabular}{ccccccc}  
    \hline  
    Model & Bandwidth(kbps) & PESQ $\uparrow$ & MelDistance $\downarrow$ & STFTDistance $\downarrow$ & SI-SDR $\uparrow$ & VISQOL(8$\sim$10s) $\uparrow$ \\  
    \hline  
    EnCodec                  & \multirow{3}{*}{1.5}    & 1.667 & 1.283       & 2.139        & -0.259 & 3.571 \\  
    $\text{EnCodec}_{ours}$  &              & 2.357 & 1.185       & 2.051        & 3.219  & 3.619 \\  
    NDVQ                     &              & \textbf{2.540} & \textbf{1.149}    & \textbf{1.991}        & \textbf{4.286}  & \textbf{3.682} \\  
    \hline  
    EnCodec                  & \multirow{3}{*}{3.0}    & 2.184 & 1.113       & 1.975        & 2.685  & 4.001 \\  
    $\text{EnCodec}_{ours}$  &              & 2.950 & 0.996       & 1.853        & 6.043  & 4.090 \\  
    NDVQ                     &              & \textbf{3.166} & \textbf{0.962}    & \textbf{1.777}        & \textbf{7.252}  & \textbf{4.117} \\  
    \hline  
    EnCodec                  & \multirow{3}{*}{6.0}        & 2.717 & 0.987       & 1.856        & 5.585  & 4.264 \\  
    $\text{EnCodec}_{ours}$  &              & 3.414 & 0.878       & 1.715        & 8.489  & 4.354 \\  
    NDVQ                     &              & \textbf{3.588} & \textbf{0.855}       & \textbf{1.661}        & \textbf{9.664}  & \textbf{4.355} \\  
    \hline  
    EnCodec                  & \multirow{3}{*}{12.0}        & 3.217 & 0.882       & 1.768        & 7.732  & 4.446 \\  
    $\text{EnCodec}_{ours}$  &             & 3.698 & 0.807       & 1.646        & 10.132 & \textbf{4.494} \\  
    NDVQ                     &             & \textbf{3.818} & \textbf{0.800}       & \textbf{1.598}        & \textbf{11.398} & 4.465 \\  
    \hline  
    EnCodec                  & \multirow{3}{*}{24.0}   & 3.556 & 0.923    & 1.725        & 8.658  & \textbf{4.540} \\  
    $\text{EnCodec}_{ours}$  &             & 3.770 & \textbf{0.790}       & 1.632        & 10.558 & 4.528 \\  
    NDVQ                     &             & \textbf{3.874} & 0.797       & \textbf{1.588}        & \textbf{12.004} & 4.477 \\  
    \hline  
    \end{tabular}  
\end{table*}  
\subsection{Training Objective}
In our training process, we combine reconstruction loss, discriminative loss, and modified codebook loss to train our NDVQ model.

\textbf{Reconstruction loss} 
To reconstruct the time and frequency domain features as much as possible, following EnCodec\cite{encodec}, the reconstruction loss is comprised of two components: time domain and frequency domain. 

For the time domain, we minimize the L1 distance between the target and reconstructed audio as
\begin{equation}
    \ell_t(x,\hat{x})=\|x-\hat{x}\|_1,
\end{equation}
where $x$ is the target audio  and $\hat{x}$ is the reconstructed audio.

For the frequency domain, we minimize the combination of L1 and L2 distance between the mel-spectrogram of the target and the reconstructed audio. Varying window lengths and hop sizes are leveraged to compute the frequency domain loss, defined as
\begin{align}
    \ell_f(x,\hat{x})=\frac{1}{|s|}\sum_{i}\|\mathbf{Mel}_i(x)-\mathbf{Mel}_i(\hat{x})\|_1 \nonumber \\ + \|\mathbf{Mel}_i(x)-\mathbf{Mel}_i(\hat{x})\|_2,
\end{align}
where $\mathbf{Mel}_i$ denotes a 64-bin mel-spectrogram with a window size selected from the set \{32, 64, 128, 256, 512, 1024, 2048\}, and a hop length chosen from \{4, 8, 16, 32, 64, 128, 256\}. $|s|$ represents the number of windows.

\textbf{Codebook Loss}
VQ models can be optimized by gradient or exponential moving average (EMA) updates \cite{vqvae}. For the former, we minimize the combination of codebook loss and commitment loss. For the latter, we use the information from historical batches to update the codebook. However, the codebook optimization by the above two methods is deterministic nearest neighbor selection, which is no longer applicable to our NDVQ method which is a distribution sample-based VQ method.

To address this issue, we modified the combination of codebook loss and commitment loss, and we also added a regularization loss function to limit the range of variance. The codebook loss and commitment loss are used to optimize the codebook's data mean component $\mu$, while the third segment incorporates L2 regularization to restrain the variance $\sigma$. This regularization is crucial as it prevents excessive variance that may introduce large noise during re-parameterization sampling. The overall codebook loss is defined as: 
\begin{equation}    
\resizebox{.91\hsize}{!}{$  
\ell_{c} = \left\| \text{sg}[\mathbf{\mu}({x})] - {z(x)} \right\|^2_2 + \beta  \left\| \mathbf{\mu}({x}) -\text{sg}[{z(x)}] \right\|^2_2 + \gamma\left\| \mathbf{\sigma}({x}) \right\|^2_2,  
$}  
\label{vq_gradient}    
\end{equation}  
where the sg denotes the stop-gradient operator, $\mu(x)$ represents the mean of the data, $\sigma(x)$ the variance, and ${z(x)}$ is the encoder output. Furthermore, $\beta$ and $\gamma$ are scalars to balance different components, with $\beta$ set to 0.25 and $\gamma$ to 0.00001.

\textbf{Discriminative Loss}
Following previous successful methods, NDVQ also employs generative adversarial training. Within this framework, the discriminator is trained to distinguish between target and reconstructed audio to enhance its reconstruction capabilities. Our discriminative loss is following EnCodec to balance both latency and audio quality.

First, the adversarial loss is defined as follows:
\begin{equation}
\ell_{a}(\hat{\boldsymbol{x}})=\frac{1}{N}\sum_{i=1}^N\max(0,1-D_{i}(\hat{\boldsymbol{x}}))),
\end{equation}
where $D_i(\hat{x})$ denotes the output of the $i$-th discriminator and $N$ is the number of sub-discriminators.

Additionally, the feature-matching loss can be formulated as follows:
\begin{equation}
    \ell_{fm}(x,\hat{x})=\frac{1}{NL}\sum_{i=1}^{N}\sum_{l=1}^{L}\frac{\|D_{i}^{l}(x)-D_{i}^{l}(\hat{x})\|_{1}}{\mathrm{E}\left[|D_{i}^{l}(x)|_{1}\right]},
\end{equation}
where $D_i^l(\hat{x})$ is the $l$-th layer feature of the $i$-th discriminator, $\hat{x}$ represents the target audio and $\hat{x}$ means the reconstructed audio and $\mathrm{E}\left[|D_{i}^{l}(x)|_{1}\right]$ represents the expected value over $i$-th layer feature.

Finally, the discriminative loss is characterized by a hinge-based adversarial loss function:
\begin{equation}
\resizebox{.91\hsize}{!}{$ 
    \ell_d(x,\hat{x})=\frac{1}{N}\sum_{i=1}^{N}\max(0,1-D_i(x))+\max(0,1+D_i(\hat{x})),
    $}
\end{equation}
where $D_i(\hat{x})$ is the $i$-th discriminator.

The overall generator loss is a weighted sum of the adversarial and the reconstruction loss, i.e.,
\begin{align}  
L_G = \lambda_t \cdot \ell_t(x,\hat{x}) + \lambda_f \cdot \ell_f(x,\hat{x}) + \lambda_a \cdot \ell_a(\hat{x}) + \nonumber \\  
\quad \lambda_{fm} \cdot \ell_{fm}(x,\hat{x}) + \lambda_c \cdot \ell_c(x),  
\end{align}  

Conversely, the discriminator is refined using its discriminative loss function:
\begin{equation}
L_D = \lambda_d \cdot \ell_d(x,\hat{x}).
\end{equation}

In our experiment, we set $\lambda_t=0.5, \lambda_f=0.5, \lambda_c=0.5, \lambda_{fm}=5,\lambda_a=1,\lambda_d=1$. Hyperparameters were selected based on a grid search, optimized for both reconstruction quality and compression efficiency.
\section{EXPERIMENTS AND RESULTS}
\label{sec:exp}
\subsection{Dataset}
To verify the efficacy of our NDVQ, we use  LibriTTS\cite{libritts} as our training dataset, which is an audio-book multi-speaker English corpus of approximately 585 hours of read English speech at a 24kHz sampling rate. During training, we randomly cropped a 1-second segment from the audio samples for each batch, and no data augmentations were used. 
\subsection{Experimental Setup}
All models were trained for 500,000 iterations, with each iteration comprising approximately 700,000 sample points (equivalent to 30 seconds within 24k Hz data). We use Adam optimizer with a learning rate of 3e-4. 
Furthermore, To prevent the discriminator from overpowering the generator in the beginning, we do not update the discriminator during the initial 50,000 updates.

\begin{table*}[ht]    
\centering    
\caption{Overview of Signal Level Metrics}    
\begin{tabular}{|p{0.13\textwidth}|p{0.75\textwidth}|p{0.08\textwidth}|}    
\hline    
\textbf{Method} & \textbf{Functionality} & \textbf{Range} \\ \hline    
PESQ $\uparrow$ & Simulates the human auditory system to objectively evaluate audio quality. & 0.5 $\sim$ 4.5 \\ \hline    
MelDistance $\downarrow$& Measures mel spectrogram perceptual differences in speech signals. & 0 $\sim$ +$\infty$ \\ \hline    
STFTDistance $\downarrow$& Quantifies Short-Time Fourier Transform spectral distance. & 0 $\sim$ +$\infty$ \\ \hline    
SI-SDR $\uparrow$& Evaluates the quality of a source sound by measuring signal-to-distortion ratio. & 0 $\sim$ +$\infty$ \\ \hline    
ViSQOL $\uparrow$& Evaluates audio quality using MOS-LQO score based on spectro-temporal similarity. & 1 $\sim$ 5 \\ \hline    
\end{tabular}    
\label{tab:signal_level_metrics}    
\end{table*}    

\subsection{Objective Evaluations}
We compared our model with the previous SOTA audio compression model EnCodec across various bandwidths, specifically {1.5, 3, 6, 12, 24}. We use the LibriTTS test-other\cite{libritts} subset as our test dataset. To evaluate the model's performance from a comprehensive perspective, we use various criteria related to perceptual quality and signal distortion. Detailed descriptions follow Table \ref{tab:signal_level_metrics}. \\
\textbf{Perceptual Quality Assessment Metrics}: To simulate the human hearing perception of reconstructed audio quality, we employed the Perceptual Evaluation of Speech Quality (PESQ)\cite{pesq} and VISQOL\cite{visqol}. Following VISQOL general guidelines, we collect 8$\sim$10s audios and resample them to 16khz. \\
\textbf{Signal Distortion Metrics}: For the quantification of signal fidelity, we applied MelDistance\cite{meldistance}, STFTDistance\cite{stftdistance}, and Scale-Invariant Signal-to-Distortion Ratio (SI-SDR)\cite{sisdr}. These metrics facilitate the comparison of the reconstructed audio against the original, providing a measure of accuracy and distortion.
\subsection{Results}
The main results are presented in Table \ref{main_table}. For a fair comparison, we reproduced the EnCodec model on the LibriTTS dataset, denoted as $\text{EnCodec}_{ours}$. The results show that the NDVQ model exhibits enhanced performance in low-bitrate transmission. In the following section, we evaluate with a fixed bandwidth of 6.0 kbps.

\subsection{Ablation Studies}
\subsubsection{Different Codebook Entropy}
As demonstrated in Table \ref{tab:codebook usage}, our method significantly improves the entropy of the codebook in the LibriTTS test-other subset. This shows that each code is used more evenly and contains richer information, indicating that our model enhances the representation space with the same codebook size.
\begin{table}[ht]
\centering
\caption{Entropy of different codebook layer, computed using code usage statistics on LibriTTS test-other}
\begin{tabular}{ccccc}
\hline
\multirow{2}{*}{Model}  & \multicolumn{4}{c}{Entropy of RVQ Layer} \\ \cline{2-5} 
                        & 1        & 8        & 16       & 32      \\ \hline
EnCodec                 & 8.21   & 8.88   & 8.88   & 8.86  \\
$\text{EnCodec}_{ours}$ & 9.38   & 9.69   & 9.60   & \textbf{9.19}  \\
NDVQ                    & \textbf{9.90}   & \textbf{9.95}   & \textbf{9.90}   & 9.10  \\ \hline
\end{tabular}
\label{tab:codebook usage}
\end{table}
\subsubsection{Inference sample method}
Introducing learnable variance and re-parameterization during training can enhance the model's robustness because it forces every code to be as far away from the codebook space. However, we believe it is not necessary during the inference stage and conduct experiments to verify our hypothesis. 
As shown in Table \ref{infer}, NDVQ achieves better reconstructed audio quality when only using the mean than re-parameterization with both mean and variance during inference. 
\begin{table}[!ht]
\centering
\caption{Comparison of Different Inference method}
\begin{tabular}{cccc}
\hline
method                     & PESQ $\uparrow$   & SI-SDR $\uparrow$ & VISQOL(8$\sim$10s) 
 $\uparrow$ \\ \hline
$\mu$                         &  \textbf{3.588} &  \textbf{9.664}      &      \textbf{4.355}   \\
$\mu + \epsilon \cdot \sigma$ &  3.436  &  9.236      &     4.296  \\ \hline
\end{tabular}
\label{infer}
\end{table}
\subsubsection{Zero-shot TTS}
Our method implicitly introduces a learnable variance and builds a safety margin for each code, encouraging the codebook representations more diverse and informative, which may be particularly advantageous for generative tasks, such as codec-based speech synthesis.

To verify this hypothesis, we trained the VALL-E model \cite{wang2023neural} on the Libriheavy dataset, applying our audio codec method trained on LibriTTS. We conducted experiments using two distinct prefix lengths for speech prompts and the results are generated with five times sampling following \cite{wang2023neural}.
We employed WavLM-TDNN \cite{chen2022wavlm} and  Hubert-Large model \cite{hsu2021hubert} which fine-tuned on the LibriSpeech 960h corpus to evaluate speaker similarity and word error rate (WER) respectively. 
As demonstrated in Table \ref{valle}, our method achieved a lower WER and higher speaker similarity in the VCTK dataset compared to $\text{EnCodec}_{ours}$ trained on the LibriTTS dataset. 
\begin{table}[htb]
\centering
\caption{Results of Zero-shot TTS}
\begin{tabular}{cccc}
\hline
Tokenizer      & Prefix  & WER$\downarrow$ & SIM$\uparrow$ \\ \hline
$\text{EnCodec}_{ours}$   & \multirow{2}{*}{3s}    & 4.1   & 0.416\\
NDVQ &    & \textbf{2.7}  & \textbf{0.426}         \\ \hline
$\text{EnCodec}_{ours}$   & \multirow{2}{*}{5s}    & 3.1   & 0.472 \\
NDVQ &         & \textbf{2.8}             & \textbf{0.482}         \\ \hline
\end{tabular}
\label{valle}
\end{table}

\subsubsection{Out-of-domain audio data compression}

The introduction of variance brings a distributed representation, which could enhance the generalization capability, and achieve better performance on out-of-domain datasets. We train NDVQ in one language and one domain LibriTTS dataset, and randomly select 1000 audios from out-of-domain datasets, the English VCTK\cite{Yamagishi2012}(in different scenario) and Chinese AISHELL-1\cite{aishell1}(in another language) dataset. 
\begin{table}[!ht]  
\centering 
\caption{Results of VCTK and AISHELL-1}
\resizebox{0.5\textwidth}{!}{
\begin{tabular}{cccccc}
\hline
Model          & \begin{tabular}[c]{@{}c@{}}Bandwidth\\ (kbps)\end{tabular} & PESQ$\uparrow$ & \begin{tabular}[c]{@{}c@{}}Mel\\ Distance$\downarrow$\end{tabular} & \begin{tabular}[c]{@{}c@{}}STFT\\ Distance$\downarrow$\end{tabular} & SI-SDR$\uparrow$ \\ \hline
\multicolumn{6}{c}{\begin{tabular}[c]{@{}c@{}}VCTK\\ (randomly select 1000 audios)\end{tabular}}  \\ \hline
$\text{EnCodec}_{ours}$ & \multirow{2}{*}{1.5}      & 1.797  & 1.134     & 1.874    & 0.855            \\
NDVQ           &               & \textbf{2.217}          & \textbf{1.054}      & \textbf{1.748}     & \textbf{1.942}\\ \hline
$\text{EnCodec}_{ours}$ & \multirow{2}{*}{3.0}  & 2.347    & 1.029      & 1.763     & 3.846            \\
NDVQ           &             & \textbf{2.854}          & \textbf{0.901}     & \textbf{1.611}    & \textbf{5.368} \\ \hline
$\text{EnCodec}_{ours}$ & \multirow{2}{*}{6.0}  & 2.870    & 0.940    & 1.679       & 6.336            \\
NDVQ           &        & \textbf{3.324}   & \textbf{0.833}    & \textbf{1.533}      & \textbf{7.919}  \\ \hline
$\text{EnCodec}_{ours}$ & \multirow{2}{*}{12.0}      & 3.127    & 0.877     & 1.626       & 7.892            \\
NDVQ           &        & \textbf{3.588}          & \textbf{0.790}     & \textbf{1.488}      & \textbf{9.710}  \\ \hline
$\text{EnCodec}_{ours}$ & \multirow{2}{*}{24.0}    & 3.183          & 0.863        & 1.614     & 8.335            \\
NDVQ           &        & \textbf{3.614}          & \textbf{0.786}       & \textbf{1.481}      & \textbf{10.445} \\ \hline
\multicolumn{6}{c}{\begin{tabular}[c]{@{}c@{}}AISHEll-1\\ (resample 24khz and randomly select 1000 audios)\end{tabular}}       \\ \hline
$\text{EnCodec}_{ours}$ & \multirow{2}{*}{1.5}  & 2.084   & 1.135     & 2.043     & 3.317            \\
NDVQ           &        & \textbf{2.208}          & \textbf{1.127}      & \textbf{1.940}        & \textbf{3.801}\\ \hline
$\text{EnCodec}_{ours}$ & \multirow{2}{*}{3.0}     & 2.664  & 0.963          & 1.790       & 6.160             \\
NDVQ           &       & \textbf{2.877}          & \textbf{0.942}  & \textbf{1.734}   & \textbf{6.995}    \\ \hline
$\text{EnCodec}_{ours}$ & \multirow{2}{*}{6.0}    & 3.265     & 0.848      & 1.673      & 8.954            \\
NDVQ           &      & \textbf{3.436}          & \textbf{0.838}   & \textbf{1.650}     & \textbf{9.761}       \\ \hline
$\text{EnCodec}_{ours}$ & \multirow{2}{*}{12.0}    & 3.655      & \textbf{0.776}       & 1.628   & 10.976           \\
NDVQ           &        & \textbf{3.807}          & 0.777    & \textbf{1.602}   & \textbf{11.855} \\ \hline
$\text{EnCodec}_{ours}$ & \multirow{2}{*}{24.0}  & 3.744          & \textbf{0.758}    & 1.608      & 11.491           \\
NDVQ           &         & \textbf{3.924}          & 0.775      & \textbf{1.599}   & \textbf{12.666} \\ \hline
\end{tabular}}
\label{vctk}
\end{table}

For a fair comparison, we employ our reproduced EnCodec $\text{EnCodec}_{ours}$ as the baseline model which is also trained on the LibriTTS dataset. 
As shown in Table \ref{vctk}, our NDVQ demonstrates significant improvements in both signal distortion and perceptual quality.
The results confirm that with the introduced variance, NDVQ can achieve much stronger generalization capability, and lead to better performance for speech reconstruction in out-of-domain scenarios.   

\section{CONCLUSION}
\label{sec:conclusion}
We introduced Normal Distribution Vector Quantization (NDVQ), which innovatively applies a distribution-based approach to vector quantization in audio codec. NDVQ's adoption of normal distributions within the codebook enhances robustness and generalization capability, leading to a better reconstructed audio quality, especially at extremely low bandwidths. Our comparative analysis with EnCodec demonstrates NDVQ's superior performance in audio compression tasks and downstream codec-based speech synthesis tasks, confirming its potential as a more resilient alternative to traditional VQ methods. While the real-world audio environment encompasses speech, ambient sounds, and music, this investigation focused solely on speech, leaving other audio domains unexplored and thus limiting the applications. The challenge of developing a universal audio compression model based on our method represents a compelling avenue for future research.
\section{ACKNOWLEDGEMENT}
\label{sec:acknowledgement}
This work was supported by the National Natural Science Foundation of China  (No. 62206171 and No. U23B2018), Shanghai Municipal Science and Technology Major Project under Grant 2021SHZDZX0102, the International Cooperation Project of PCL.

\bibliographystyle{IEEEbib}
\bibliography{refs}

\end{document}